\documentclass[12pt]{article}

\usepackage{amsfonts,amsmath}
\usepackage{latexsym}

\usepackage{epsfig}
\usepackage{psfrag}

\usepackage{epsfig}
\usepackage[T1]{fontenc} 
\usepackage[utf8]{inputenc} 
\usepackage[icelandic,english]{babel}

\def\void{}
\def\labelmark{\marginpar{\small\labelname}}

\newenvironment{formula}[1]{\def\labelname{#1}
\ifx\void\labelname\def\junk{\begin{displaymath}}
\else\def\junk{\begin{equation}\label{\labelname}}\fi\junk}%
{\ifx\void\labelname\def\junk{\end{displaymath}}
\else\def\junk{\end{equation}}\fi\junk}

{\ifx\void\labelname\def\junk{\end{array}\end{displaymath}}
\else\def\junk{\end{array}\right.\end{equation}}
\fi\junk\labelmark\def\labelname{}\def\junk{}
}

\newcommand{\beq}{\begin{formula}}
\newcommand{\eeq}{\end{formula}}
\newcommand{\beqv}{\begin{formula}{}}

\newcommand{\rf}[1]{(\ref{#1})}

\newcommand{\bea}{\begin{eqnarray}}
\newcommand{\eea}{\end{eqnarray}}
\newcommand{\beas}{\begin{eqnarray*}}
\newcommand{\eeas}{\end{eqnarray*}}
\newcommand{\beqs}{\begin{displaymath}}
\newcommand{\eeqs}{\end{displaymath}}

\newcommand{\ben}{\begin{equation}}
\newcommand{\een}{\end{equation}}

\newcommand{\bdm}{\begin{displaymath}}
\newcommand{\edm}{\end{displaymath}}

\newcommand{\N}{{\cal N}}

\begin{document}

\selectlanguage{english}

 \topmargin 0pt
 \oddsidemargin 5mm
 \headheight 0pt
 \topskip 0mm

\addtolength{\baselineskip}{0.5\baselineskip}

\pagestyle{empty}

\hfill

\vspace{1cm}

\begin{center}

{\Large \bf Appearance of vertices of infinite order \\
in a model of random trees}

\medskip
\vspace{1.5 truecm} 

{\large \bf \today} 

\vspace{1.5 truecm}

{\bf Thordur Jonsson and Sigurður Örn Stefánsson}

\vspace{0.4 truecm}

The Science Institute, University of Iceland

Dunhaga 3, 107 Reykjavik

Iceland

 \vspace{.7 truecm}

 \vspace{1.3 truecm}

\end{center}

\noindent {\bf Abstract.} We study an equilibrium statistical
mechanical model of tree graphs which are made up of a linear subgraph
(the spine) to which leaves are attached.   We prove that 
the model has two phases, a generic
phase where the spine becomes infinitely long in the
thermodynamic limit and all vertices have finite order 
and a condensed phase where the spine is finite with probability one
and a single vertex of infinite order appears in the thermodynamic limit.
We calculate the spectral dimension of the graphs in both phases 
and prove the
existence of a Gibbs measure.  We discuss generalizations of this model 
and the relationship with models of nongeneric random trees.

 \newpage
 \pagestyle{plain}

\section{Introduction}  The study of random graphs has been an active
area of research in mathematics and physics for the past few decades
and remains so.  In particular, the study of random trees and random
triangulations has found many applications in theoretical physics, see
e.g.\ \cite{book}.  Our understanding of the equilibrium statistical
mechanics of trees with local action is fairly good but not
complete.  By local action we mean 
 an action which is given by a sum
 over the vertices and only depends on their order.
It is now known that so called generic trees can be viewed as critical Galton-Watson processes \cite{GT} 
which are very well understood mathematically \cite{GW}. 
A corresponding picture has not been fully established for 
nongeneric trees which are more difficult to analyse.  
Much of our knowledge about
such trees comes from numerical simulations and educated guesswork 
\cite{bbbj,bb1,bbj,bck,exotic}.
However, a consistent picture has emerged \cite{MS}.  Typically a 
vertex of infinite order
appears in the thermodynamic limit 
but full analytic control of this phase of random trees is still missing.

In this paper we study a simple model of random graphs which
exhibits the same behavior as random trees with a local action, 
namely there is a generic
phase where the free energy can be calculated by a saddle point
technique and a nongeneric phase where a vertex of infinite order
appears in the thermodynamic limit.  This model was analysed
extensively some years ago in a series of papers \cite{bbbj,bb1,bbj}
under the name ``balls in boxes'' and ``backgammon'' model.  
Closely related models appear in the study of the equilibrium distribution
for urn models and zero range processes, see e.g.\ \cite{evanshanney,godreche} and
references therein.

The graphs that underlie the model studied in this paper
have been called caterpillar graphs or simply
caterpillars by
graph theorists \cite{cat} and we will adopt that name here. 
Caterpillars
are defined as graphs with the property that all vertices of order
higher than one form a linear subraph, i.e.\ if all leaves are removed
one ends up with a linear graph.  Various applications of caterpillar
graphs in physics and chemistry are described in \cite{acat}.

When the caterpillar grows
large two things can happen: it either becomes very long or some of
the vertices will have a large number of leaves.  A priori these two
phenomena could coexist but we will see that this is not the case in
the model we consider.  Our main motivation is to study the
appearance of a vertex of infinite order in a rigorous fashion.

This paper is organized as follows.  In the next section we define the
model, establish our notation and derive some simple properties.  In
section 3 we study the generic phase and prove that
generic caterpillars are infinitely long in the thermodynamic limit 
with all vertices of finite order.  We calculate the order
distribution explicitly.  The Hausdorff and spectral dimensions of
generic caterpillars are both shown to be equal to 1.  In section 4,
which is the core of this paper, 
we study nongeneric caterpillars
and begin by establishing an asymptotic formula for the canonical
partition function.  We then prove that there arises 
exactly one vertex of infinite
order in the thermodynamic limit.  We find the probability distribution
of the distance from the root of the random caterpillar (taken to be
one of the endpoints of the spine) to the infinite
order vertex as well as the probability distribution for the orders of
the other vertices.  

The nongeneric caterpillar graphs have infinite 
Hausdorff and spectral dimensions since there is
a vertex of infinite order at a finite distance from the root with
probability one.  However, we will show that the spectral dimension
defined in terms of the ensemble average of the return probability of
random walker is finite and varies continuously with the parameters of
the model.  

In section 5 we comment on generalizations of
this model and discuss nongeneric trees and how they are related to
the caterpillar model.  In an appendix we establish the existence of a
probability measure on the set of infinite caterpillar graphs where
vertices may have infinite order.

\section{The model} 
\begin {figure} [h]
\centerline{\scalebox{1}{\includegraphics{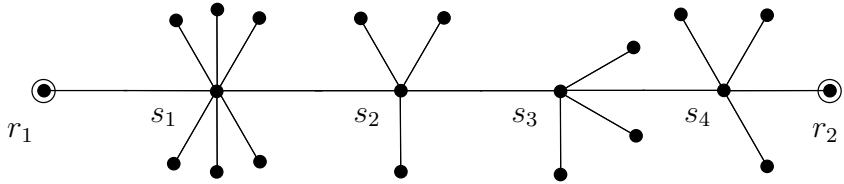}}}
\caption{An example of a finite caterpillar graph.} 
\label{f:barb}
\end {figure}

A finite caterpillar is a finite graph which consists of a
linear graph, which we call the spine, to which leaves (i.e.~individual links) are attached.  We mark the end vertices of the linear graph
by $r_1$ and $r_2$ and call $r_1$ the root of the caterpillar.  
Both these vertices have order one by definition.  
Furthermore, we will view the caterpillars as planar graphs so we distinguish
between left leaves and right leaves, see Fig.~\ref{f:barb}.  The assumption 
of planarity is not essential.
We denote the set of all caterpillars
with $N$ edges by $B_N$. For a caterpillar $\tau \in B_N$, denote the graph distance between
$r_1$ and $r_2$ by $\ell(\tau)$ and call it the length of the caterpillar. For a caterpillar of length $\ell$ we denote the vertices on the spine between $r_1$ and $r_2$ by $s_1,\ldots,s_{\ell-1}$. 

Let $w_n$, $n=1,2,\ldots $, 
be a sequence of nonnegative numbers which will be called
weight factors.  The weight of a caterpillar $ \tau \in B_N$ is defined as
\begin {equation}
 w(\tau) = \prod_{i \in \tau \setminus \{r_1,r_2\}} w_{\sigma(i)}~,
\end {equation}
where $\sigma(i)$ denotes the order of the vertex $i$ and by abuse of
notation we let $\tau$
also denote the set of vertices in $\tau$.
We define the finite volume partition function by
\begin {equation} \label{fvpf}
Z_N = \sum_{\tau \in B_N} w(\tau)
\end {equation}
and a probability distribution on $B_N$ by
\begin {equation} \label{prob}
 \nu_N(\tau) = \frac{w(\tau)}{Z_N}.
 \end {equation}
 The weight factors $w_n$, or alternatively the measures $\nu_N$, define
 what we call a caterpillar ensemble. 

Since the probability of a given caterpillar only depends on the order of its vertices, an equivalent way of defining this ensemble is the following. If $\tau \in B_N$ consider the finite sequence $c(\tau) = \left(\sigma(s_1),\sigma(s_2),\ldots,\sigma(s_{\ell-1})\right)$ and assign to it the probability
\begin {equation} \label{redef}
\tilde{\nu}_N(c(\tau)) = \nu_N(\tau) \prod_{i=1}^{\ell(\tau)-1}(\sigma(s_i)-1).
\end {equation}
The product factor in (\ref{redef}) accounts for the number of different caterpillars which correspond to the same sequence $c(\tau)$. Define the set $\tilde{B}_N = \{ c(\tau) ~|~ \tau \in B_N\}$. It is clear that $(B_N,\nu_N)$ is equivalent to $(\tilde{B}_N,\tilde{\nu}_N)$ in the sense that $\nu_N(\tau)$ only depends on $c(\tau)$. This allows us to extend the notion of finite caterpillars to infinite ones:
\begin {equation} \label{generalcaterpillars}
 \tilde{B} = \left\{ \big(b_i\big)_{i=1}^{k-2} ~|~ k, b_i \in \{2,3,\ldots\}\cup\{\infty\}, 1 \leq i \leq k-2 \right\}
\end {equation}
where $k=2$ corresponds to the unique caterpillar of length $\ell = 1$. Note that an element in $\tilde{B}$ which has infinite terms and/or infinite length has no counterpart in $B_N$ for any $N$.

\begin {figure} [h]
\centerline{\scalebox{1}{\includegraphics{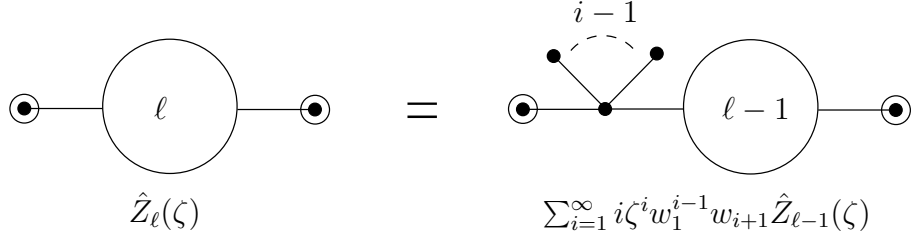}}}
\caption{An illustration of the recursion (\ref{e:rec}).} 
\label{f:rec}
\end {figure}

Define the finite volume partition
function with fixed distance $\ell$ between $r_1$ and $r_2$ as
\begin {equation}
Z_{N,\ell} = \sum_{\tau \in B_N, \ell(\tau) = \ell} w(\tau).
\end {equation}
It is useful to work with the generating functions
\begin {equation} \label{gffv}
 Z(\zeta) = \sum_{N = 1}^\infty Z_N \zeta^N
 \end {equation}
 and
 \begin {equation} \label{gfunction}
 g(z) = \sum_{n=0}^{\infty} w_{n+1}z^n
 \end {equation}
 with radii of convergence $\zeta_0$ and $\rho$, respectively, both of
 which we assume to be nonzero.
 Define also
 \begin {equation}
  \hat{Z}_\ell(\zeta) = \sum_{N=1}^{\infty}Z_{N,\ell} \zeta^N.
  \end {equation}
  Then it is clear that
  \begin {equation} \label{zl}
  Z(\zeta) = \sum_{\ell=1}^{\infty} \hat{Z}_\ell(\zeta).
  \end {equation}
We have the recursion relation
\begin {equation}\label{e:rec}
\hat{Z}_\ell(\zeta) = \zeta g'(w_1\zeta) \hat{Z}_{\ell-1}(\zeta),
\end {equation}
for any $\ell\geq 1$, see Fig.~\ref{f:rec}. 
Using the above equation and $\hat{Z}_1(\zeta) = 
\zeta$  gives 
\begin {equation} \label{ltree}
\hat{Z}_\ell(\zeta) = \zeta \Big(\zeta g'(w_1 \zeta)\Big)^{\ell-1}
\end {equation}
and by \rf{zl}
\begin {equation} \label{treeformula}
Z(\zeta) = \frac{\zeta}{1-\zeta g'(w_1 \zeta)}.
\end {equation}
From \rf{treeformula} we see that 
$\zeta_0$ is the smallest solution 
of the equation
\begin {equation} \label{genconditionI}
 \zeta g'(w_1\zeta) = 1
 \end {equation}
on the interval $(0 , \rho/w_1)$ if such a solution exists. If it does not 
exist then $\zeta_0 = \rho/w_1$.

If $\zeta_0 < \rho/w_1$ then $g$ is analytic at $w_1\zeta_0$ and we say 
that we have a {\it generic ensemble}.  This has been called the ``fluid
phase'' by other authors \cite{bbj}.  If $\zeta_0 = \rho/w_1$ we have a 
{\it nongeneric ensemble}. Notice that if $\rho = \infty$ then
the ensemble is always generic.  For nongeneric ensembles we 
therefore have finite $\rho$. In that case we can always choose 
$\rho=1$ by scaling the weights $w_n \rightarrow w_n \rho^{n-1}$. This 
scaling does not affect the probabilities (\ref{prob}).

Now consider weights factors with $\rho = 1$ and let $w_1$ be a free 
parameter. The genericity condition is then $\frac{1}{w_1}g'(1) > 1$, 
i.e.\ $w_1 < w_c$ where
\begin {equation}
w_c \equiv g'(1) = \sum_{n=2}^{\infty} (n-1)w_n
\end {equation}
is a critical value for $w_1$. If $w_1 = w_c$ we have a nongeneric 
ensemble which we refer to as {\it critical} and if $w_1 > w_c$ we have a 
nongeneric ensemble which we refer to as {\it subcritical}.  This phase
has been called the ``condensed phase'' in the literature \cite{bbj}. 

\section{The generic phase}
Let $w_n$ be weight factors with $w_1 \neq 0$
and $w_n \neq 0$ for some $n>2$ which lead to a generic ensemble.

\noindent
{\bf Lemma 1.}
~{\it Under the stated assumptions on the weight factors, the asymptotic
behaviour of $Z_N$ is given by
\begin {equation} \label{ZNgeneric}
 Z_N = {1 \over  g'(w_1\zeta_0)+\zeta_0 w_1 g''(w_1\zeta_0)}
 \zeta_0^{-N}
  (1 + O(N^{-1}))
  \end {equation}
  if the integers $n>0$ for which $w_{n+1} \neq 0$ have no common
  divisors greater than 1. Otherwise, if their greatest common divisor
  is $d \geq 2$, then
  \begin {equation}
  Z_N = {d \over  g'(w_1\zeta_0)+\zeta_0 w_1 g''(w_1\zeta_0)}
  \zeta_0^{-N}
  (1 + O(N^{-1}))
  \end {equation}
  if $N = 1$ mod $d$, and $Z_N = 0$ otherwise.
}
\bigskip

  The proof of this Lemma is standard, cf.\ \cite{flajolet}, 
  where the corresponding
  result for generic trees is established. For generic caterpillars one can show by a straightforward application
of the methods of \cite{bergfinnur} (see also Appendix A) that the
measures $\tilde{\nu}_N$ converge as $N\to\infty$ to a measure $\tilde{\nu}$
which is concentrated on locally finite caterpillars of infinite length and the orders of the vertices on the infinite spine are independently and identically distributed by

\beq{deg}
\phi(n) = \zeta_0 (n-1)w_{n}(w_1\zeta_0)^{n-2}, \quad\quad n\geq 2.
\eeq

Denote the expectation with respect to the measure $\tilde{\nu}$ by
$\langle\cdot\rangle_{\tilde{\nu}}$.  If $V_r$ is the number of vertices within
a distance $r$ from the root the Hausdorff dimension $d_H$ is defined as
\beq{haus}
\langle V_r\rangle_{\tilde{\nu}}\sim r^{d_H}.
\eeq
We write $f(x)\sim x^{\gamma}$ if for any $\epsilon >0$ there are
constants $C_1$ and $C_2$ such that $C_1x^{\gamma +\epsilon}\leq f(x)\leq 
C_2x^{\gamma -\epsilon}$. If $\langle V_r\rangle_{\tilde{\nu}}$ increases faster than any power of $r$ then we say that $d_H$ is infinite. We see from \rf{deg} that the expectation value \rf{haus} is 
\beq{dist}
\langle V_r \rangle_{\tilde{\nu}} = (\zeta_0g''(w_1\zeta_0)-1)(r-1) +1.
\eeq
It follows that the Hausdorff dimension of generic caterpillars is $1$.

Let $p_\tau (t)$ be the probability that a simple random walk which
leaves the root of an infinite 
caterpillar $\tau$ at time 0 is back at the root at time
$t$, i.e.\ after $t$ steps.  If there exists a number $d_s>0$ such that 
\beq{spec}
p_\tau (t)\sim t^{-d_s/2}
\eeq
as $t\to\infty$ then we say that the spectral dimension of the graph
is $d_s$. If $p_\tau(t)$ decays faster than any power of $t$ then we say that $d_s$ is infinite.  For a discussion of the spectral dimension of some random
graph ensembles, see \cite{DJW,GT,SDRB}.   

The spectral dimension is most
coveniently analysed by generating functions.  We define
\beq{qdef}
Q_\tau(x)=\sum_{t=0}^\infty p_\tau (t)(1-x)^{t/2}
\eeq
and let $Q(x)=\langle Q_\tau(x)\rangle_{\tilde{\nu}}$. 
We define $p_\tau^{(1)}(t)$ to be the probability that a 
simple random walk which
leaves the root at time $0$ is back at the root for the first time
after $t$ steps and let $P_\tau (x)$ be the corresponding generating
function defined as $Q_\tau (x)$ with $p_\tau (t)$ replaced by 
$p_\tau^{(1)}(t)$.  Then we have the relation
\beq{relation}
Q_\tau (x)={1\over 1-P_\tau (x)}.
\eeq
Let $n$ be the smallest nonnegative integer for which $Q^{(n)}_\tau(x)$ diverges as $x \rightarrow 0$. If
\begin {equation}
(-1)^nQ_\tau^{(n)}(x)\sim x^{-\alpha} 
\end {equation}
for some $\alpha\in [0,1)$ then clearly 
\beq{specalpha}
d_s=2(1-\alpha+n),
\eeq
if $d_s$ exists. We define the spectral dimension of the caterpillar ensemble by \rf{specalpha} provided $(-1)^{n} Q^{(n)}(x)\sim
x^{-\alpha}$.

From the monotonicity lemmas in \cite{SDRB} we get an upper 
bound $x^{-1/2}$ on 
$Q(x)$ by throwing away all the legs of the caterpillar.
To get a lower bound on $Q(x)$ we use a slight modification of Lemma 7 
in \cite{GT} which is the following.  For a given 
infinitely long caterpillar $\tau$ with a first return probability 
generating function $P_\tau(x)$ we have, for all integers $L\geq1$ 
and $0<x\leq 1$,
\beq{yyy}
P_\tau(x) \geq 1-\frac{1}{L} - x \sum_{i=1}^L \sigma(s_i(\tau)).
\eeq
We then get, using \rf{relation}, \rf{yyy} and Jensen's inequality,
\beq{886}
Q(x) \geq \frac{1}{1-\langle P_\tau(x) \rangle_{\tilde{\nu}}} 
\geq \frac{1}{\frac{1}{L}+\langle \sigma(s_1)\rangle_{\tilde{\nu}} Lx}.
\eeq
In the generic phase we see from equation 
(\ref{deg}) that $\langle \sigma(s_1)\rangle_{\tilde{\nu}}$ is finite. 
Choosing $L = \left[x^{-1/2}\right]$ we find

\beq{885}
Q(x) \geq c x^{-1/2}
\eeq
where $c$ is a constant. 
It follows from \rf{885}, the upper bound on $Q(x)$ and \rf{specalpha} that
the spectral dimension of generic caterpillars is $d_s = 1$.

\section{The subcritical phase}
In this section we begin by calculating the asymptotic behaviour of the
canonical partition function in the subcritical phase.  We then show
that there is exactly one vertex of infinite order in the
thermodynamic limit. The mechanism leading to a unique vertex of infinite order is similar to the one leading to a unique spine for generic trees \cite{bergfinnur,GT}.  We calculate the probability distribution for
the location of the infinite order vertex as well as the probability
distribution for the orders of the other vertices.  Finally, we
discuss the spectral dimension of subcritical caterpillars.

\begin {figure} [h]
\centerline{\scalebox{0.4}{\includegraphics{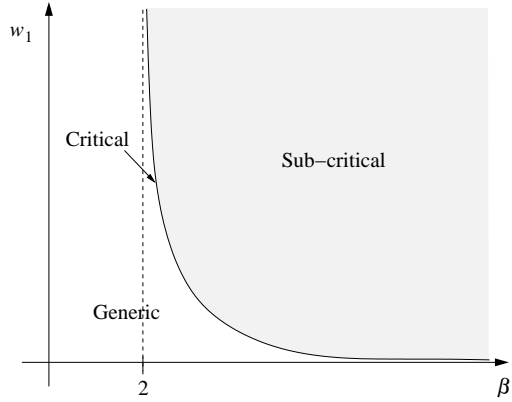}}}
\caption{A diagram showing the different phases of the caterpillars.}
\label{d_phase}
\end {figure}

We take $\rho=1$ and $w_1 > w_c$ so that we are in the
subcritical 
phase.  We study a concrete model where
\beq{weights}
w_i = i^{-\beta}, \quad \quad i \geq 2,
\eeq
and let $w_1$ be a free parameter in the specified range. We comment on extensions in Section 5.
Figure \ref{d_phase} shows the phase diagram of the caterpillars.
A necessary condition for being in the subcritical phase is 
$\beta>2$ since otherwise $w_c=\infty$.

\medskip
\noindent
{\bf Lemma 2.} {\it
For the weights given in 
(\ref{weights}) and $w_1>w_c$ we have
\begin {equation} \label{ZNnongeneric}
Z_N = \frac{1}{(w_1-w_c)^2} N^{1-\beta} w_1^N\big(1+o(1)\big)
\end{equation}
as $N\to\infty$.}

\medskip
\noindent
{\it Proof.} We can write
\begin {equation}
 Z_N = \sum_{\ell=1}^N Z_{N,\ell}.
 \end {equation}
Define a sequence of functions $f_N$ on the positive integers by
\beq{deff}
 f_N(\ell) = \left\{ \begin{array}{ll}
w_1^{-N}N^{\beta-1}Z_{N,\ell} &							
\quad\textrm{$\ell \leq N$}\\
0 & \quad\textrm{$\ell>N$.}\\
\end{array} \right.
\eeq
We claim that
\beq{claim}
\lim_{N\rightarrow\infty}f_N(\ell) =
\frac{1}{w_c^2}(\ell-1)\left(\frac{w_c}{w_1}
\right)^\ell \equiv f(\ell).
\eeq
We accept the claim for a moment and finish the proof of the Lemma.

It is clear that $f_N(\ell)$ is summable for every $N$. We also see that $f(\ell)$ is summable since $w_1 > w_c$. Note that
for $\ell\leq N$
\begin {eqnarray}
f_N(\ell) & = & w_1^{-\ell} N^{\beta-1} \sum_{N_1+\ldots+N_{\ell-1} = N-\ell}
\prod_{i=1}^{\ell-1}\left\lbrace (N_i+1)w_{N_i+2}\right\rbrace \nonumber\\
&\leq& w_1^{-\ell} N^{\beta-1} (\ell-1)
\sum_{\substack{N_1+\ldots+N_{\ell-1} =
N-\ell\\ N_1 \geq \frac{N-\ell}{\ell-1}}} \frac{N_1+1}{(N_1+2)^\beta}
\prod_{i=2}^{\ell-1}\left\lbrace
(N_i+1)w_{N_i+2}\right\rbrace \nonumber\\
&\leq& \frac{1}{w_c^2}\left(\frac{w_c}{w_1}\right)^\ell
\frac{N^{\beta-1}(N-1)}
{\left(\frac{N-l}{\ell-1}+2\right)^\beta} \leq C
(\ell-1)^\beta \left(\frac{w_c}{w_1}
\right)^\ell \label{star}
\end {eqnarray}
where $C$ is a positive constant. The first inequality in \rf{star} is
obtained by observing that at least one of the indices $N_i$ must be
larger than ${N- \ell\over \ell-1}$ and in the second one we used the
definition of $w_c$.   It follows that the sequence
$\{ f_N\}_1^{\infty}$ is
dominated by a summable function and we can calculate the limit
\beq{limitII}
 \lim_{N\rightarrow\infty}\left(w_1^{-N}N^{\beta-1}Z_N\right) =
  \lim_{N\rightarrow\infty}\sum_{\ell=1}^{\infty}f_N(\ell) =
  \sum_{\ell=1}^{\infty}f(\ell) =
   \frac{1}{\left(w_1-w_c\right)^2}.
   \eeq
This implies the desired result.  

It remains to prove the claim \rf{claim}.  
There is at least one index $i$ in the sum defining
$f_N(\ell)$ such that
$N_i \geq \frac{N-\ell}{\ell-1}.$ If there is another index $j\neq i$ such
that
$N_j > A$ where $A>1$ is a constant then we get an upper bound on that
contribution to $f_N(\ell)$ of the form
\begin {eqnarray}
 &&   w_1^{-\ell} N^{\beta -1} (\ell-1)^2  
  \sum_{\substack{N_1 + \ldots + N_{l-1} = N-\ell \\ N_1 \geq
  \frac{N-\ell}{\ell-1}\nonumber 
   \\ N_2 > A}\quad}
   \frac{N_1+1}{\left(N_1+2\right)^\beta}\prod_{i=2}^{\ell-1}
    \left\lbrace (N_i+1)w_{N_i+2}\right\rbrace \\
    &\leq& C(\ell) \frac{N^\beta }{\left(N+\ell-2\right)^\beta}
    \sum_{N_3,\ldots,N_{\ell-1}\geq 0}
    \prod_{i=3}^{\ell-1}\left\lbrace (N_i+1)w_{N_i+2}\right\rbrace
    \sum_{N_2>A}(N_2+1)
    w_{N_2+2} \nonumber\\
    &\leq& D(\ell) w_c^{\ell-3} \sum_{N_2>A}(N_2+a)w_{N_2+2}\label{plus}
    \end {eqnarray}
    where $C(\ell)$ and $D(\ell)$ are numbers which only depend on $\ell$. The
    last 
    expression goes to zero as $A \rightarrow \infty$ since $g'(1)$ is
    finite. 
   The remaining contribution to $f_N(\ell)$ is
    \begin {eqnarray*}
    &&  w_1^{-\ell} N^{\beta -1}(\ell-1)
    \sum_{\substack{N_1 + \ldots N_{\ell-1} 
    = N-\ell\\   N_1 \geq
      \frac{N-\ell}{\ell-1}
      \\N_j \leq A, \quad j\neq 1}}\prod_{i=1}^{\ell-1}\left\lbrace
    (N_i+1)w_{N_i+2}
    \right\rbrace \\
    &\substack{\longrightarrow\\N\rightarrow\infty}& 
    w_1^{-\ell} (\ell-1)\left(\sum_{n=0}^{A}(n+1)w_{n+2}\right)^{\ell-2} \\
    &\substack{\longrightarrow\\A\rightarrow\infty}&
    w_c^{-2}
    (\ell-1)\left(\frac{w_c}{w_1}\right)^\ell.
    \end {eqnarray*}
This completes the proof.
\begin{flushright} $\square$
\end{flushright}
\medskip

From the above lemma we obtain the following result.

\medskip
\noindent
{\bf Theorem 1.} {\it
For the weight factors given in (\ref{weights}) with $w_1 > w_c$
the probability  that the distance between $r_1$ and $r_2$ is $\ell$ as the 
caterpillar size $N$ goes to infinity is given by

\begin {equation} \label{lengthd}
\psi(\ell) \equiv \lim_{N\rightarrow \infty}\frac{Z_{N,\ell}}{Z_N} = 
(\ell-1)\left(1-\frac{w_1}{w_c}\right)^2\Big(\frac{w_c}{w_1}\Big)^{\ell}.
\end {equation}
For a given $\ell$, exactly one of the vertices on the spine has an infinite order, and the 
orders of the other vertices 
are identically and independently distributed by
\begin {equation} \label{phidistribution}
\phi(k) = \frac{1}{w_c}(k-1)k^{-\beta}, \quad\quad\quad k \geq 2.
\end {equation}
}

\medskip
\noindent
{\it Proof.} Combining Lemma 2 with \rf{claim} we obtain 
\rf{lengthd}.  If the length of an infinite caterpillar is $\ell<\infty$
it is clear that there is one or more vertices of infinite order.  The
inequality \rf{plus} shows that there can be at most one vertex of
infinite order in the limit $N\to\infty$.   Finally, the distribution of
the orders of the vertices which have a 
finite order in the thermodynamic limit is obtained by an
argument similar to the one leading to Equation \rf{deg}, cf. Equation (\ref{limit}). 
\begin{flushright} $\square$
\end{flushright}
\medskip

In the appendix we prove the existence of a measure $\tilde{\nu}$ on the set of infinite caterpillars which describes the subcritical phase and is obtained as the limit of the finite volume measures. The above theorem then implies that the Hausdorff dimension
$d_H$ of a random caterpillar in the subcritical phase is almost sureley (a.s.) infinite
since with probability one there is a ball of finite radius which contains infinitely many
vertices.  Similarly, the spectral dimension is a.s.~infinite because a random walk which hits the infinite order vertex
returns to the root with probability $0$. From the analysis below one can easily check that the return probability on a randomly chosen subcritical caterpillar $\tau$, $p_\tau(t)$, decays faster than any power of $t$.

In the remainder of this section we show how the definition of the spectral dimension in terms of the ensemble average with respect to $\tilde{\nu}$, see (\ref{specalpha}), leads to a spectral dimension
\begin {equation} \label{specensemble}
 d_s = 2(\beta-1)
\end {equation}
in the subcritical phase. We will refer to the unique vertex of infinite order as the ``trap''.   If the walk 
hits the trap it returns to the 
root with probability zero. Therefore, the part of the caterpillar
beyond the trap 
is irrelevant for the random walk.  When finding the spectral dimension it is 
therefore natural to consider the probability that the trap is at a distance $\ell$ 
from the root instead of considering the probability of the total length of the 
caterpillar given in (\ref{lengthd}). 

For a caterpillar of a given length, all the vertices between $r_1$ and 
$r_2$ are 
equally likely to be of infinite order so the probability 
that the trap is at a distance 
$\ell$ from root is given by
\begin {equation}
p(\ell) = \sum_{k=\ell+1}^{\infty}\frac{\psi(k)}{k-1} = \Big(1-\frac{w_c}{w_1}\Big)\Big(
\frac{w_c}{w_1}\Big)^{\ell-1} .
\end {equation}
From now on we will disregard the part of the caterpillar 
beyond the trap. Let $B_{\ell,k}$ be the set of caterpillars with distance $\ell$ between root and trap and which 
have one vertex of order $k$ and all other 
vertices of order no greater than $k$, with the exception of the
trap of course.  Let $a(k)$ be the probability that a given vertex on the spine between the root and the trap has order no greater than $k$. Then
\begin {equation}
 a(k) = \sum_{q=2}^k \phi(q).
\end {equation}
The probability that at least one of these vertices 
has order $k$ and all the others 
have order no greater than $k$ is then
\beq{probc}
 c(k,\ell) = a(k)^{\ell-1} - a(k-1)^{\ell-1}.
\eeq
Let $\pi_{\ell,k}(\tau)$ be the $\tilde{\nu}$--probability of the 
caterpillar $\tau \in B_{\ell,k}$ given 
that we are selecting from $B_{\ell,k}$.
The average return generating 
function for the subcritical caterpillars is then
\begin{equation} \label{frgf}
Q(x) = \sum_{\ell=1}^{\infty}p(\ell) \sum_{k=2}^\infty c(k,\ell) 
\sum_{\tau \in B_{\ell,k}}
\pi_{\ell,k}(\tau)Q_\tau(x).
\end{equation}

For a given distance $\ell$ between root and trap we denote by $M_\ell$ the linear subgraph which 
starts at the root and ends at the trap.
\begin {figure} [h]
\centerline{\scalebox{1}{\includegraphics{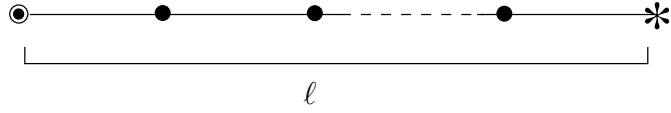}}}
\caption{The graph $M_\ell$. The root is denoted by a circled vertex and the trap by 
an asterisk.}
\end {figure}
The first return generating function for $M_\ell$ is given by
\begin {equation} \label{frgfml}
 P_{M_\ell}(x) = 1-\sqrt{x}\frac{(1+\sqrt{x})^\ell+(1-\sqrt{x})^\ell}{(1+\sqrt{x})^\ell
 -(1-\sqrt{x})^\ell},
\end {equation}
see e.g.\ \cite{DJW}.
Now attach $k$ links to each vertex of the graph $M_\ell$  except the root and the trap and denote the resulting 
graph by $M_{\ell,k}$. Using the methods of \cite{SDRB} we find that 
the first return generating 
function for $M_{\ell,k}$ is
\begin {equation} \label{renorm}
P_{M_{\ell,k}}(x) = \left(1+\frac{k}{2}x\right)P_{M_\ell}(x_k(x))
\end {equation}
where
\begin {equation}
 x_k(x) = \frac{\frac{k^2}{4}x^2+(1+k)x}{\left(1+\frac{k}{2}x\right)^2}.
\end {equation}

\begin {figure} [h]
\centerline{\scalebox{1}{\includegraphics{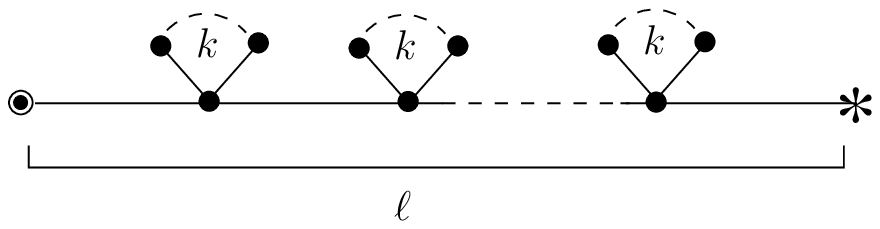}}}
\caption{The graph $M_{\ell,k}$.}
\end {figure}

To find an upper bound on the spectral dimension of subcritical caterpillars we establish a lower bound on the $n$-th derivative of the average return generating function.  Let $n$ be the smallest positive integer such 
that $Q^{(n)}(x)$ diverges as $x\rightarrow 0$. We see in the following 
calculations that we have to choose $n$ such that $n+1 < \beta \leq n+2$.  By (\ref{relation}) we find that $(-1)^n Q^{(n)}_{\tau} \geq (-1)^n P^{(n)}_{\tau}$ for any $\tau$. Thus, by differentiating (\ref{frgf}) $n$ times and throwing away every term in the sum over $\ell$  except $\ell=2$ we get the lower bound  
\begin {equation}
 (-1)^n Q^{(n)}(x) \geq (-1)^{n}\left(1-\frac{w_c}{w_1}\right)
 \frac{w_c}{w_1} \sum_{k=2}^\infty \phi(k) P^{(n)}_{M_{2,k-2}}(x).
\end {equation}
 We easily find that
\begin {equation}
 P_{M_{2,k-2}}(x) = \frac{1-x}{2+(k-2)x}
\end {equation}
and show by induction that
\begin {equation} \label {PM2Kd}
 P^{(n)}_{M_{2,k-2}}(x) = (-1)^{n}n!\frac{(k-2)^{n-1}k}{(2+(k-2)x)^{n+1}}.
\end {equation}
Then, by (\ref{phidistribution}) and (\ref{PM2Kd}), 
\begin {eqnarray} \nonumber
(-1)^{n} \sum_{k=2}^\infty \phi(k) P^{(n)}_{M_{2,k-2}}(x)  
&=& \frac{n!}{w_c}\sum_{k=2}^\infty \frac{(k-2)^{n-1}k^{1-\beta}
(k-1)}{(2+(k-2)x)^{n+1}} \\ \label {aint}
&\geq& C x^{\beta-n-2} \int_x^{\infty}\frac{y^{n+1-\beta}}{(2+y)^{n+1}}dy	
\end {eqnarray}
where $C>0$ is a constant. If $\beta < n + 2$ the last integral is convergent 
when $x\rightarrow 0$ but if $\beta = n+2$ it diverges logarithmically. In 
both cases we get an upper bound for the spectral dimension $d_s \leq 2(\beta-1)$.

To find a lower bound on the spectral dimension of subcritical
caterpillars we establish an upper bound on the $n$-th derivative of the average return 
generating function.  First note that $1 > a(k) = a(k-1) + \phi(k)$ and therefore
\begin {eqnarray}
c(k,\ell) &=& \left(a(k)-a(k-1)\right) \nonumber\\ 
&\times& \left(a(k)^{\ell-2}+a(k)^{\ell-3}a(k-1)+\ldots + 
a(k)a(k-1)^{\ell-3}+a(k-1)^{\ell-2}\right) \nonumber\\ \label{probest}
&\leq& \phi(k)(\ell-1).
\end {eqnarray}

Now consider a caterpillar $\tau \in B_{\ell,k}$ and the graph $M_\ell$. Denote the vertices on the spine of $M_\ell$ between the root and the trap by $s_1,s_2,\ldots,s_{\ell-1}$. One can obtain the graph $\tau$ from $M_{\ell}$ by attaching $m_\tau(s_i)$ links to $s_i$, $i=1,\ldots,\ell-1$ where $0 \leq m_\tau(s_i) \leq k-2$. Using the methods of \cite{SDRB} we can write
\begin {equation} \label{qtau}
 Q_{\tau}(x) = \sum_{\substack{\omega:~ r_1 \rightarrow r_1~\\ \textrm{on}~M_\ell}} K_\tau(x,\omega)W_{M_\ell}(\omega) (1-x)^{|\omega|/2}
\end {equation}
where the sum is over all random walks $\omega$ on $M_\ell$ which begin and end at the root,
\begin {equation}
 K_\tau(x,\omega) = \prod_{\substack{t=1 \\ \omega_t \in \{s_1,\ldots,s_{\ell-1}\}}}^{|\omega| -1} \left(1+\frac{m_\tau(\omega_t)}{2}x\right)^{-1},
\end {equation}
\begin {equation}
 W_{M_\ell}(\omega) = \prod_{t=0}^{|\omega|-1} (\sigma(\omega_t))^{-1}
\end {equation}
where $\omega_t$ is the vertex at which $\omega$ is located at step $t$ and $|\omega|$ denotes the length of $\omega$. The $i$--th derivative of the function $K_\tau(x,\omega)$ can be estimated as
\begin {equation} \label{kderest}
(-1)^i \frac{d^i}{dx^i}K(x,\omega) \leq H(|\omega|) \frac{(k-2)^i}{(2+(k-2)x)^i}
\end {equation}
where $H$ is a polynomial with positive coefficients. From the relation (\ref{relation}) and the explicit formula (\ref{frgfml}) one can easily see that $(-1)^{i}Q^{(i)}_{M_\ell}(0)$ is a positive polynomial in $\ell$ of degree $2i+1$. Therefore, differentiating (\ref{qtau}) $n$ times and using the estimate (\ref{kderest}) we get the upper bound
\begin {equation} \label{tauest}
(-1)^n Q_\tau^{(n)}(x) \leq \sum_{i=0}^{n} S_i(\ell) \frac{(k-2)^i}{(2+(k-2)x)^i}
\end {equation}
where the $S_i$ are positive polynomials in $\ell$.
Differentiating (\ref{frgf}) $n$ times w.r.t.~$x$ and using the estimates (\ref{probest}) and (\ref{tauest}) we finally obtain
\begin {equation}
 (-1)^n Q^{(n)}(x) \leq \sum_{i=0}^{n} \sum_{\ell=1}^{\infty}
p(\ell) S_i(\ell)(\ell-1) \sum_{k=2}^\infty \phi(k) 
\frac{(k-2)^i}{(2+(k-2)x)^{i}}.
\end {equation}
The sum over $\ell$ is convergent since $S_i$ is a polynomial in $\ell$ and $p(\ell)$ decays exponentially. The sum over $k$ is estimated from above by an integral as in (\ref{aint}) which yields a lower bound on the spectral dimension $d_s \geq 2(\beta -1)$. This proves \rf{specensemble}.

\section{Discussion}
In this paper we have given a description of the phases of the random caterpillar model. However, it is not complete. First of all, in the subcritical nongeneric phase, when $w_1 > w_c$ we limit ourselves to the particular choice of weights in \rf{weights}. This strict power law can easily be relaxed to an asymptotic power law. It is however not clear how to generalize this to arbitrary weights satisfying $w_1 > w_c$.

Secondly, we have no rigorous results on what happens on the critical line of the phase diagram in Fig.~\ref{d_phase} when $w_1 = w_c$. This problem is discussed in similar models in \cite{bbj,bck} where it is argued that when $g''(1)<\infty$ the phase is characterised as the generic phase and when $g''(1) = \infty$ the critical exponent of $Z_N$ changes continuously with $\beta$. 

The order of the phase transition from the condensed phase to the fluid phase also depends on whether $g''(1)$ is finite or infinite. Define the free energy as
\begin {equation}
F(w_1) = \lim_{N\rightarrow \infty}\frac{\log Z_N(w_1)}{N}.
\end {equation}
Using (\ref{genconditionI}), (\ref{ZNgeneric}) and (\ref{ZNnongeneric}) one finds that
\begin {equation}
F'(w_1) = \left\{
 \begin{array}{cl}
 \left(\frac{1}{\zeta_0^2g''(w_1\zeta_0(w_1))}+w_1\right)^{-1} & \quad \text{if } \quad w_1 < w_c\\
   w_1^{-1} & \quad \text{if } \quad w_1 > w_c
 \end{array} \right.
\end {equation}
and thus 
\begin {equation}
\lim_{w_1\rightarrow w_c ^-} F'(w_1)= \frac{1}{\frac{w_c^2}{g''(1)}+w_c}.
\end {equation}
This shows that when $g''(1) < \infty$ the phase transition is first order but when $g''(1) = \infty$ it is continuos in agreement with \cite{bbj,bck}.

The caterpillar model can be generalized to more complicated tree models by replacing the leaves on the spine by trees with vertices of order bounded by $K$, the caterpillars corresponding to $K = 1$. With similar analysis as for the caterpillars, one obtains two phases: a fluid phase (generic) and a condensed phase (nongeneric), seperated by a critical value of $w_1$ given by
\begin {equation}
 w_c(K) = g'(1) - \sum_{n=2}^K w_n.
\end {equation}
In the fluid phase, the finite volume probability measures converge to a measure concentrated on trees with an infinite spine with critical Galton Watson outgrowths analogous to the generic trees in \cite{GT}. In the crumpled phase the measures converge to trees with spine of a finite length $\ell$ distributed by 
\begin {equation}
 \psi(\ell,K) = (\ell-1)\left(1-\frac{w_1}{w_c(K)}\right)^2\left(\frac{w_c(K)}{w_1}\right)^\ell.
\end {equation}
Exactly one of the vertices on the spine has infinite degree and the order of other vertices is independently distributed by 
\begin {equation}
 \phi(k,K) = \frac{1}{w_c(K)}(k-1)w_k, \quad \quad k \geq 2.
\end {equation}
The outgrowths from the spine are independent subcritical Galton Watson trees with offspring probabilities
\begin {equation}
 p_n(K) = \frac{w_{n+1}}{\sum_{n=1}^K w_n},\quad \quad 0 \leq n \leq K-1.
\end {equation}
As $N \rightarrow \infty$ one finds that the size of the large vertex is approximately $(1-m(K))N$ where $m(K)<1$ is the mean offspring probability of the Galton Watson process. This is in agreement with analogous results in \cite{bbj,bck,MS}. What makes the calculations easy in the condensed phase in the above models is the fact that the large vertex which emerges as $N\rightarrow \infty$ has to stay on the spine due to the restriction on the order of the vertices in the outgrowths. When the cutoff on the vertex orders is removed ($K=\infty$) one obtains nongeneric trees. In this case it is more difficult to locate the large vertex and one has to use other methods in the calculations.  However, we expect the above characterisation of the condensed phase to hold with minor adjustements as is argued in \cite{MS}. This will be addressed in a forthcoming paper on nongeneric trees.

\medskip

\noindent {\bf Acknowledgment.}  This work is supported in part by
Marie Curie grant MRTN-CT-2004-005616, the Icelandic Science Fund, the University of Iceland Research Fund and the Eimskip Research
Fund at the University of Iceland. We would like to acknowledge hospitality at the Jagellonian University and discussions
with Piotr Bialas, Zdzislaw Burda, Bergfinnur Durhuus and Jerzy Jurkiewicz.

\section*{Appendix A - The Gibbs measure in the condensed phase}

\bigskip
In this appendix we consider the set $\tilde{B}$ of all caterpillars defined in (\ref{generalcaterpillars}).  We equip this set with a metric and
adopt the methods of \cite{bergfinnur} (see also \cite{angel,billingsley}) 
to prove the existence of a probability measure
on this set which describes the subcritical phase.

We define a metric $d$ on $\tilde{B}$ by 
\beq{metric}
d(b,c) = \left\{\begin {array}{cc}\max
\left\{\frac{1}{1+\min\{b_i,c_i\}}~\Big|~ b_i \neq c_i\right\} & \text{if $\ell(b) = \ell(c)$,}\\
1 & \text{otherwise}
\end {array}\right.
\eeq
where $b = (b_1,b_2,\ldots)$ and $c = (c_1,c_2,\ldots)$. We define the maximum of the empty set to be 0. It is an elementary calculation to verify that this definition
fulfills the axioms for a metric.
Denote the open ball centered at $b$ and with radius $s$ by
$\mathcal{B}_s(b)$. It is easy to verify that these balls are both open and closed and that if $c \in \mathcal{B}_s(b)$ 
then $\mathcal{B}_s(c) = \mathcal{B}_s(b)$.  Denote the set of caterpillars of fixed length $\ell$ by $\tilde{B}^{(\ell)}$. For any $\ell\in\mathbb{N}$ the set $\tilde{B}^{(\ell)}$ is compact. Define
\begin {equation}
 \tilde{B}' = \bigcup_{N=1}^\infty \tilde{B}_N.
\end {equation}
The set $\tilde{B}'$ is a countable dense subset of $\tilde{B}$.

From now on we consider the weight factors (\ref{weights}) with $w_1 > w_c$. The probability measures $\tilde{\nu}_N$ on $\tilde{B}_N$ will be shown to converge to a measure $\tilde{\nu}$ on $\tilde{B}$. 

\medskip
\noindent
{\bf Theorem A1.} {\it
For the weight factors (\ref{weights}) with $w_1 > w_c$ the measures 
$\tilde{\nu}_N$ viewed as probability measures on $\tilde{B}$ converge 
weakly to a measure ${\tilde{\nu}}$ as $N \rightarrow \infty$ 
and ${\tilde{\nu}}$ is concentrated on the set of caterpillars of finite length with exactly 
one vertex of infinite order. The length of the spine is distributed by (\ref{lengthd}). All the vertices between $r_1$ and $r_2$ are equally likely to be of infinite order and the orders of the others are independently distributed by (\ref{phidistribution}).
}



\medskip
\noindent
{\it Proof.}
Applying the methods of \cite{bergfinnur} 
we need to show the following:

\begin {enumerate}
 \item The sequence $\left( {\tilde{\nu}}_N\left(\mathcal{B}_{\frac{1}{k}}\left(b\right)\right)\right)_{N=1}^\infty$ converges for all $k \in \mathbb{N}$ and all $b \in \tilde{B}'$.
 
\item For every $\epsilon > 0$ there exists a compact subset $C \subseteq \tilde{B}$ such that

\begin {equation}
 {\tilde{\nu}}_N\left(\tilde{B}\setminus C\right) < \epsilon, \quad\quad 
 \text{for all $N \in \mathbb{N}$.}
\end {equation}
\end {enumerate}

To prove Property 1 take a finite caterpillar $b = (b_1,\ldots,b_{\ell(b)-1})\in\tilde{B}'$. In order to
streamline the notation we write $\ell(b) = \ell$. Denote the set of 
indices $i$ for which $b_i < k$ by $\underline{I}$ and the set of 
indices $i$ for which $b_i \geq k$ by $\overline{I}$.   Then
\begin {equation}
\mathcal{B}_{\frac{1}{k}}\left(b\right) = \left\lbrace c \in \tilde{B}^{(\ell)}~|~ c_i = b_i ~\text{if}~ i\in\underline{I},
 ~ ~c_i \geq k ~\text{if}~ i\in\overline{I}  \right\rbrace. 
\end {equation}
Denote the number of elements in $\overline{I}$ by $R$. Now order the 
indices in $\overline{I}$ in increasing order and for a given caterpillar 
in $\mathcal{B}_{\frac{1}{k}}\left(b\right)$ let $N_i, ~ 1\leq i \leq R$ be the term in the caterpillar 
corresponding to the $i$-th index in $\overline{I}$. 
We can then write
\begin {eqnarray} \label{Aprob}
{\tilde{\nu}}_N\left(\mathcal{B}_{\frac{1}{k}}\left(b\right)\right) &=& Z_N^{-1} w_1^{N-\ell} W_0 \sum_{\substack{N_1+\ldots+N_R=N+\ell-2-b_0 
\\ N_i \geq k, ~\forall i}} ~\prod_{i=1}^R\left[\left( N_i-1)w_{N_i}
\right)\right]  
\end {eqnarray}
where 
\begin {equation*}
b_0 = \sum_{i\in\underline{I}}b_i \quad\quad\quad 
\text{and}\quad\quad\quad W_0 = \prod_{i\in\underline{I}}
\left[\left(b_i-1\right)w_{b_i}\right].
\end {equation*}
First note that if $\overline{I}$ is empty then ${\tilde{\nu}}_N\left(\mathcal{B}_{\frac{1}{k}}\left(b\right)\right) 
\longrightarrow 0$ when $N \longrightarrow \infty$. If it is not 
empty, there exists an index $i\in \overline{I}$ in the above sum 
such that $N_i \geq \frac{N+l-2-b_0}{R}$. If there is another 
index $j\neq i$ such that $N_j > C$ where $C\geq k$ is a constant 
then we get an upper bound 
\begin {equation}
K \sum_{N_2 > C}\left(N_2-1\right)w_{N_2}
\end {equation}
on that contribution to the above sum using (\ref{ZNnongeneric}) and the methods in the proof of Lemma 2 where $K$ is a positive number which only depends on $b$ and $k$. 
The last expression goes to zero as $C\longrightarrow \infty$ since 
$g'(1)$ is finite. Estimating the remaining contribution to 
(\ref{Aprob}) we get
\begin {eqnarray} \nonumber
&&(w_1-w_c)^2 w_1^{-l} N ^{\beta-1} W_0 \sum_{i=1}^R~
\sum_{\substack{N_1+\ldots+N_R=N+l-2-b_0 \\ k \leq N_j \leq C, 
\quad j \neq i}} ~\prod_{i\in\overline{I}}\left[\left( N_i-1)w_{N_i}
\right)\right]\left(1+o(1)\right) \\ \nonumber
&\substack{\longrightarrow \\ N\rightarrow \infty}& 
(w_1-w_c)^2 w_1^{-l}  W_0 R \left(\sum_{n=k}^{C} (n-1)w_{n}\right)^{R-1} 
\\ \label{limit}
&\substack{\longrightarrow \\ C \rightarrow \infty}& 
(w_1-w_c)^2 w_1^{-l}  W_0 R \left(\sum_{n=k}^{\infty} (n-1)w_{n}\right)^{R-1}
\end {eqnarray}
proving the convergence. The calculations  show that the 
measure is concentrated on the set of caterpillars with 
exactly one infinite term.

In order to prove Property 2 we take our compact set to be
\begin {equation}
C_L = \bigcup_{\ell=1}^L \tilde{B}^{(\ell)}
\end {equation}
and we need to show that 
\begin {equation}
{\tilde{\nu}}_N\left( \left\lbrace b \in \tilde{B} ~\Big|~ \ell(b) > L 
\right\rbrace \right) \longrightarrow 0 \quad\quad 
\text{as} \quad \quad L \longrightarrow \infty
\end {equation}
uniformly in $N$. We estimate as in the proof of Lemma 2
\begin {eqnarray*}
{\tilde{\nu}}_N\left( \left\lbrace b \in \tilde{B}~|~ \ell(b) = \ell 
\right\rbrace \right) &=&  \frac{Z_{N,l}}{Z_N} \leq C \left(\frac{w_c}{w_1}\right)^l  (l-1)^\beta
\end {eqnarray*}
where $C$ is a constant. Since $w_1 > w_c$ this completes the proof of
the convergence. The distribution of the length of the spine and order of vertices follows from (\ref{limit}).
\begin{flushright} $\square$
\end{flushright}
\medskip


\end{document}